\documentclass[fleqn,twoside]{article}
\usepackage{latexsym}
\usepackage{amssymb,amsmath}

\usepackage[headings]{espcrc2}

\newcommand{\refl}[1]{(\ref{#1})}
\newcommand{\eeql}[1]{\label{#1}\eeq}
\newcommand{\vp}{\ensuremath{\phi}}
\newcommand{\lag}{\ensuremath{\mathcal{L}}}

\newcommand{\ra}{\ensuremath{\rightarrow}}

\newcommand{\zpr}{\ensuremath{Z'}}
\newcommand{\mzp}{\ensuremath{M_{Z'}}}

\newcommand{\uprm}{\ensuremath{U(1)'}}

\newcommand{\skipblk}[1]{}                                                      
\def\bqa{\begin{eqnarray}}                                                      
\def\eqa{\end{eqnarray}}

\newcommand{\douba}[2]{\ensuremath{                                                  
\left( \begin{array}{c} #1    \\ #2 
        \end{array}\right)}}

\newcommand{\st}{\ensuremath{SU(2)}}

\newcommand{\x}{\ensuremath{\times}}

\newcommand{\sinn}{\ensuremath{\sin^2\theta_W\,}}

\newcommand{\beq}{\begin{equation}}                                             
\newcommand{\eeq}{\end{equation}}     
                                                                                   
\newcommand{\oh}{\ensuremath{\frac{1}{2}}}

\usepackage{graphicx}
\title{\zpr\ Physics at the LHC}

\author{Paul Langacker\address{Institute for Advanced Study
,Princeton, NJ 08540}\thanks{This work was supported by  NSF PHY-0503584 and by the IBM Einstein Fellowship.}}

\usepackage{ifthen}
\newboolean{hyperlinks}
\setboolean{hyperlinks}{false}
\setboolean{hyperlinks}{true}
\ifthenelse{\boolean{hyperlinks}}{\usepackage{hyperref}\bibliographystyle{h2-elsevier} 
}{\newcommand{\href}[2]{#2}
\newcommand{\url}[1]{}\bibliographystyle{h2-elsevier} }
\begin{document}

\begin{abstract}
The existing limits on \zpr\ gauge bosons and prospects for discovery and diagnostic studies at the LHC are briefly reviewed.\vspace{1pc}
\end{abstract}

\maketitle

\section{Introduction}\label{ints}
Additional \zpr\ gauge bosons  occur frequently in extensions of the 
standard model (SM) or its minimal supersymmetric extension (MSSM), usually emerging as
an unbroken ``remnant''  of a larger gauge symmetry. Examples include superstring constructions, 
grand unified theories, extended electroweak groups, or alternatives to the minimal Higgs mechanism
for electroweak breaking. Kaluza-Klein excitations  of the SM gauge bosons also occur in models involving large and/or warped extra dimensions provided the gauge bosons are free to propagate in the bulk, with 
$M\sim R^{-1}\sim 2 {\ \rm TeV }\x (10^{-17} {\rm  cm }/R)$ in the large dimension case.
The new \zpr s may occur at any mass scale, but here we concentrate on  the
TeV scale relevant to the LHC, which is especially motivated by
supersymmetric \uprm\ models (in which both the electroweak and \uprm\ breaking scales are usually
set by the soft supersymmetry breaking parameters) and by alternative models of electroweak symmetry breaking. We first briefly review the formalism and the existing constraints from precision electroweak
(weak neutral current, $Z$-pole, LEP 2, and FCNC) measurements and from direct searches at the Tevatron, and then comment on the prospects for a \zpr\ discovery, diagnostics of its
couplings, and related issues such as the associated extended Higgs and neutralino sectors at the LHC. Much more extensive discussions of specific models and other implications, along with a more
complete set of references, are given in several reviews~\cite{Langacker:2008yv,Rizzo:2006nw,Leike:1998wr}. Other recent developments, especially the possibility of a \zpr\ as a ``portal''
to a quasi-hidden sector, such as may be associated with dark matter or supersymmetry breaking, 
were reviewed in~\cite{Langacker:2009im,Goodsell:2009xc}.

\section{Formalism}\label{formalism}
The interactions of the photon ($A$), $Z$ (i.e., $Z^0_1$) and other flavor-diagonal neutral gauge bosons
with fermions  is
\beq -\lag_{NC}= \underbrace{eJ^\mu_{em} A_\mu+g_1 J^\mu_1 Z^0_{1 \mu}}_{SM}+ \sum_{\alpha=2}^{n+1} g_\alpha J^\mu_\alpha Z^0_{\alpha \mu}, \eeql{f1}
where $g_\alpha$ are the gauge couplings  (with $g_1=g/\cos\theta_W$), and the currents are
\beq J^\mu_\alpha= \sum_i \bar f_i \gamma^\mu[\epsilon_L^{{\alpha}}(i)P_L+\epsilon_R^{{\alpha}}(i)P_R] f_i. \eeql{f2}
 $\epsilon_{L,R}^{{\alpha}}(i)$ are the $U(1)_\alpha$  charges of the left- and right-handed components of  fermion $f_i$, and the theory is chiral for $\epsilon_{L}^{{\alpha}}(i)\ne \epsilon_{R}^{{\alpha}}(i)$. 
We also define the vector and axial couplings
 \beq g_{V,A}^{{\alpha}}(i) =\epsilon_L^{{\alpha}}(i)\pm\epsilon_R^{{\alpha}}(i). \eeql{f1a}
 It is often convenient to instead
 specify the charges $Q_\alpha$ for  the left-chiral 
fermion $f_L$ {and}  and left-chiral antifermion  $f^c_L$,
\beq Q_{\alpha f}=\epsilon_L^{{\alpha}}(f) \quad  \qquad Q_{\alpha f^c}=-\epsilon_R^{{\alpha}}(f). \eeql{f3}
For example, the SM charges for the $u_L$ and $u^c_L$ are
$Q_{1u}=\oh - \frac{2}{3} \sinn$ and
$Q_{1u^c}=+ \frac{2}{3} \sinn$.
One can similarly define the $U(1)_\alpha$  charge of the scalar field $\vp$  as $Q_{\alpha\vp}$.

For a single extra \zpr, the $Z-\zpr$ mass matrix after symmetry breaking is
\beq M^2_{Z-Z'} =   \left(
\begin{array}{cc}
 M_{Z^0}^2 & \Delta^2\vspace*{2pt} \\  \Delta^2 & M_{Z'}^2
\end{array} \right). \eeql{f4}
If, for example, the symmetry breaking is due to an \st-singlet $S$ and two doublets
$\phi_u= \douba{\phi^0_u}{\phi^-_u}, \  \phi_d= \douba{\phi^+_d}{\phi^0_d}$,
then
\beq
\begin{split}
M_{Z^0}^2 =&  \frac{1}{4} g_1^2 \bigl(|\nu_u|^2+|\nu_d|^2\bigr)  \\
 \Delta^2 = & \oh g_1 g_2 \bigl(Q_u |\nu_u|^2-Q_d |\nu_d|^2 \bigr) \\
M_{Z'}^2 =& g_2^2 \bigl( Q_u^2 |\nu_u|^2 + Q_d^2|\nu_d|^2+Q_S^2 |s|^2 \bigr),
\end{split}
\eeql{f5}
where
\beq \begin{split}
 \nu_{u,d} &\equiv \sqrt{2} \langle  \phi^0_{u,d} \rangle, \qquad s=\sqrt{2}  \langle S \rangle\\
\nu^2&=|\nu_u|^2+|\nu_d|^2\sim (246 {\rm\ GeV})^2 .
\end{split}
\eeql{f6}
The physical mass eigenvalues are $M_{1,2}^2$, the physical gauge particles are $Z_{1,2}$, and the  mixing angle
 $\theta_{Z\zpr}$ is given by
 $ \tan^2\theta_{Z\zpr} ={(M_{Z^0}^2-M_1^2)}/{(M_2^2-M_{Z^0}^2)}$.
In the important special case  $M_{Z'} \gg (M_{Z^0}, |\Delta|)$ one finds
\beq \begin{split}
M_1^2 & \sim M_{Z^0}^2 - \frac{\Delta^4}{M_{Z'}^2}\ll M_2^2, \qquad M_2^2 \sim M_{Z'}^2 \\
\theta_{Z\zpr} & \sim -\frac{\Delta^2}{M_{Z'}^2}\sim C \frac{g_2}{g_1} \frac{M_1^2}{M_2^2}
\\ C& \equiv
2\left[  \frac{Q_u |\nu_u|^2-Q_d |\nu_d|^2}{|\nu_u|^2+|\nu_d|^2} \right].
\end{split} \eeql{f8}

A \uprm\ can yield a natural solution to the supersymmetric $\mu$ problem~\cite{Kim:1983dt}
(unless the charges are obtained from $B-L$ and $Y$), by forbidding
an elementary $\mu$ term but allowing the superpotential term $W \sim \lambda_S S H_u H_d$, where $S$ is a 
SM singlet charged under the \uprm. This induces an effective $\mu$ parameter $ \mu_{eff} = \lambda_S \langle S \rangle$,
which is usually of the same scale as the soft supersymmetry breaking parameters~\cite{Suematsu:1994qm,Cvetic:1995rj,Cvetic:1997ky}. This mechanism is similar to the NMSSM (e.g.,~\cite{Accomando:2006ga,Maniatis:2009re}), but is automatically free of induced tadpole and domain wall problems.

We have so far implicitly assumed canonical kinetic energy terms for the $U(1)$ gauge bosons. However,  $U(1)$ gauge invariance allows a more general kinetic mixing~\cite{Holdom:1985ag},
\beq\begin{split}
 \lag_{kin} \ra & -\frac{1}{4} F^{0\mu\nu}_1 F^0_{1\mu\nu}  -\frac{1}{4} F^{0\mu\nu}_2 F^0_{2\mu\nu}
\\  &\qquad -\frac {\sin \chi}{2} F^{0\mu\nu}_1 F^0_{2\mu\nu}
\end{split}\eeql{f9}
for $U(1)_1\x U(1)_2$.
Such terms are usually absent initially, but a (usually small) $\chi$ may be induced by loops, e.g., from nondegenerate heavy particles, in running couplings
if heavy particles decouple, or at the string level. The kinetic terms may be put in canonical form by the non-unitary transformation
\beq  \douba{Z^0_{1\mu}}{Z^0_{2 \mu}}= 
\left(
\begin{array}{cc} 1 &   -\tan \chi   \\ 0 &   1/\cos \chi\end{array}\right)
\douba{\hat Z^0_{1\mu}}{\hat Z^0_{2\mu}},
\eeql{f10}
where  the $\hat Z^0_{\alpha}$  may still undergo ordinary mass mixing, as in \refl{f4}.
The kinetic mixing has a negligible effect on masses for  $|M_{Z_1}^2| \ll |M_{Z_2}^2|$
and $|\chi|\ll 1$, but the current coupling to the heavier boson is shifted,
\beq  -\lag \ra  g_1 J^\mu_1\hat Z_{1 \mu}+  (g_2 J^\mu_2 -g_1 \chi J^\mu_1)\hat Z_{2 \mu} .
\eeql{f11}

The \zpr\ mass and mixing may also be generated by the St\"uckelberg
mechanism~\cite{Stueckelberg:1900zz,Kors:2004dx,Feldman:2006ce,Nath:2008ch}.

\section{Existing Limits}\label{existing}
\zpr s with electroweak coupling are mainly  constrained by precision electroweak data,
direct searches at the Tevatron, and searches for flavor changing neutral currents (FCNC).
 Low energy weak neutral current
 processes, which are still very important,  would be affected by  $Z_2$ exchange and by $Z-\zpr$ mixing~\cite{Erler:2009jh,Durkin:1985ev,Amaldi:1987fu,Costa:1987qp,Langacker:1991pg,Amsler:2008zzb}.
The effective four-Fermi WNC interaction becomes
\beq
 -\lag_{eff}=\frac{4G_F}{\sqrt{2}} (\rho_{eff} J^2_1+2wJ_1J_2+yJ^2_2), \eeql{a1}
where
\beq \begin{split} \rho_{eff}=&\rho_1 \cos^2 \theta_{Z\zpr} + \rho_2 \sin^2 \theta_{Z\zpr} \\
w=&\frac{g_2}{g_1}\cos\theta_{Z\zpr}\sin\theta_{Z\zpr} (\rho_1-\rho_2) \\
y =&\left(  \frac{g_2}{g_1} \right)^2 (\rho_1 \sin^2 \theta_{Z\zpr} + \rho_2 \cos^2 \theta_{Z\zpr}),
\end{split}
\eeql{a2}
with
\beq
\rho_\alpha\equiv M_W^2/(M^2_\alpha \cos^2 \theta_W).
\eeql{a2a}
 The $Z$-pole experiments at LEP and  SLC~\cite{lep:2005ema} are extremely sensitive to $Z-\zpr$ mixing, which 
  shifts $M_1$ downward from the SM expectation and also affects the $Z_1$ vector and axial
 vertices, which become 
\beq \begin{split}
V_i =&\cos \theta_{Z\zpr}  g^1_V(i) + \frac{g_2}{g_1} \sin \theta_{Z\zpr}   g^2_V(i) \\
A_i =&\cos \theta_{Z\zpr}  g^1_A(i) + \frac{g_2}{g_1} \sin \theta_{Z\zpr}   g^2_A(i) .
\end{split}
\eeql{a3}
However, the $Z$-pole experiments have little sensitivity
 to $Z_2$ exchange. At LEP2~\cite{Alcaraz:2006mx} virtual $Z_2$ exchange leads to a four-fermi operator, analogous to the
 $\rho_2$ part of $\lag_{eff}$ in \refl{a1}, which interferes with the $\gamma$ and $Z$.

\begin{figure}[htb]
\includegraphics*[scale=0.41]{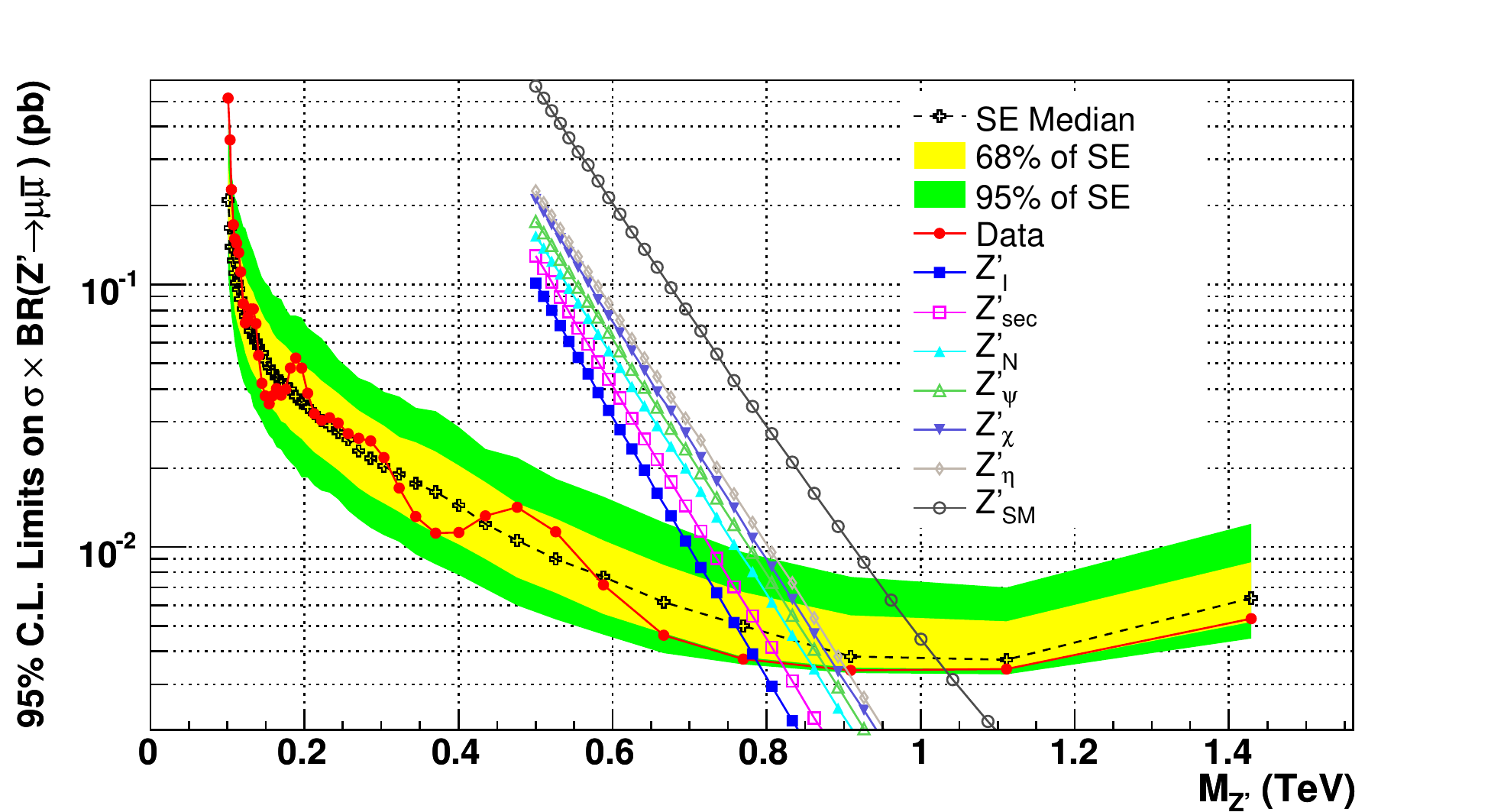}
\caption{CDF limits on various \zpr\ models from the dimuon channel, from \cite{Aaltonen:2008ah}.}
\label{CDFmuons}
\end{figure}

The CDF~\cite{Aaltonen:2008ah,Aaltonen:2008vx} and D\O~\cite{do:2009} collaborations at the Tevatron have 
searched for Drell-Yan resonances, especially  $ \bar p p \ra e^+ e^-, \mu^+ \mu^-$~\cite{Robinett:1981yz},
as illustrated in Figure \ref{CDFmuons}.
 In the narrow width approximation,
the tree-level rapidity distribution for $AB\ra Z_\alpha$ is
\beq \begin{split}
\frac{d\sigma_{\zpr}}{dy}=& \frac{4\pi^2 x_1x_2}{3M_\alpha^3} \sum_{i} \bigl(f_{q_i}^A(x_1)f_{\bar q_i}^B(x_2)\bigr. \\
&+\bigl. f_{\bar q_i}^A(x_1)f_{q_i}^B(x_2)\bigr)
\Gamma (Z_\alpha \ra q_i \bar q_i),
\end{split}
\eeql{a4}
where the $f$'s are the parton distribution functions, the partial widths are
\beq \Gamma (Z_\alpha \ra f_i \bar f_i)= \frac{g_\alpha^2 C_{f_i} M_\alpha}{24\pi} 
\bigl( \epsilon_L^{{\alpha}}(i)^2+\epsilon_R^{{\alpha}}(i)^2 \bigr),
\eeql{a5}
$x_{1,2}=(M_\alpha/\sqrt{s}) e^{\pm y}$, and $ C_{f_i}$ is the color factor.
More detailed estimates for the Tevatron and LHC are given in 
\cite{Leike:1998wr,Amsler:2008zzb,Weiglein:2004hn,Dittmar:2003ir,Carena:2004xs,Kang:2004bz,Fuks:2007gk,Petriello:2008zr,Baur:2001ze,Papaefstathiou:2009sr,Diener:2009vq},
including discussions of parton distribution functions, higher order QCD and electroweak effects, fermion mass corrections, decays into bosons or Majorana fermions,  width effects, 
resolutions, and backgrounds.
%: figure  cdf limits

Other search  channels relevant to hadron colliders include $\zpr \ra e^\pm\mu^\mp$~\cite{Abulencia:2006xm}; $\tau^+\tau^-$~\cite{Acosta:2005ij}; $jj$, where $j=$ jet~\cite{Weiglein:2004hn,Aaltonen:2008dn}; $b\bar b$; and  $t\bar t$~\cite{Han:2004zh,cdf:2007dz,Baur:2008uv}.
Another important probe is
the forward-backward asymmetry for  $p p (\bar p p) \ra\ell^+\ell^-$
(as a function of rapidity, $y$, for $pp$) due to $\gamma,Z,\zpr$ interference below the \zpr\
peak~\cite{Langacker:1984dc,Rosner:1995ft,Dittmar:1996my,Abulencia:2006iv}.

\begin{figure*}[htb!]
\begin{minipage}[t]{5cm} 
\includegraphics*[scale=0.34]{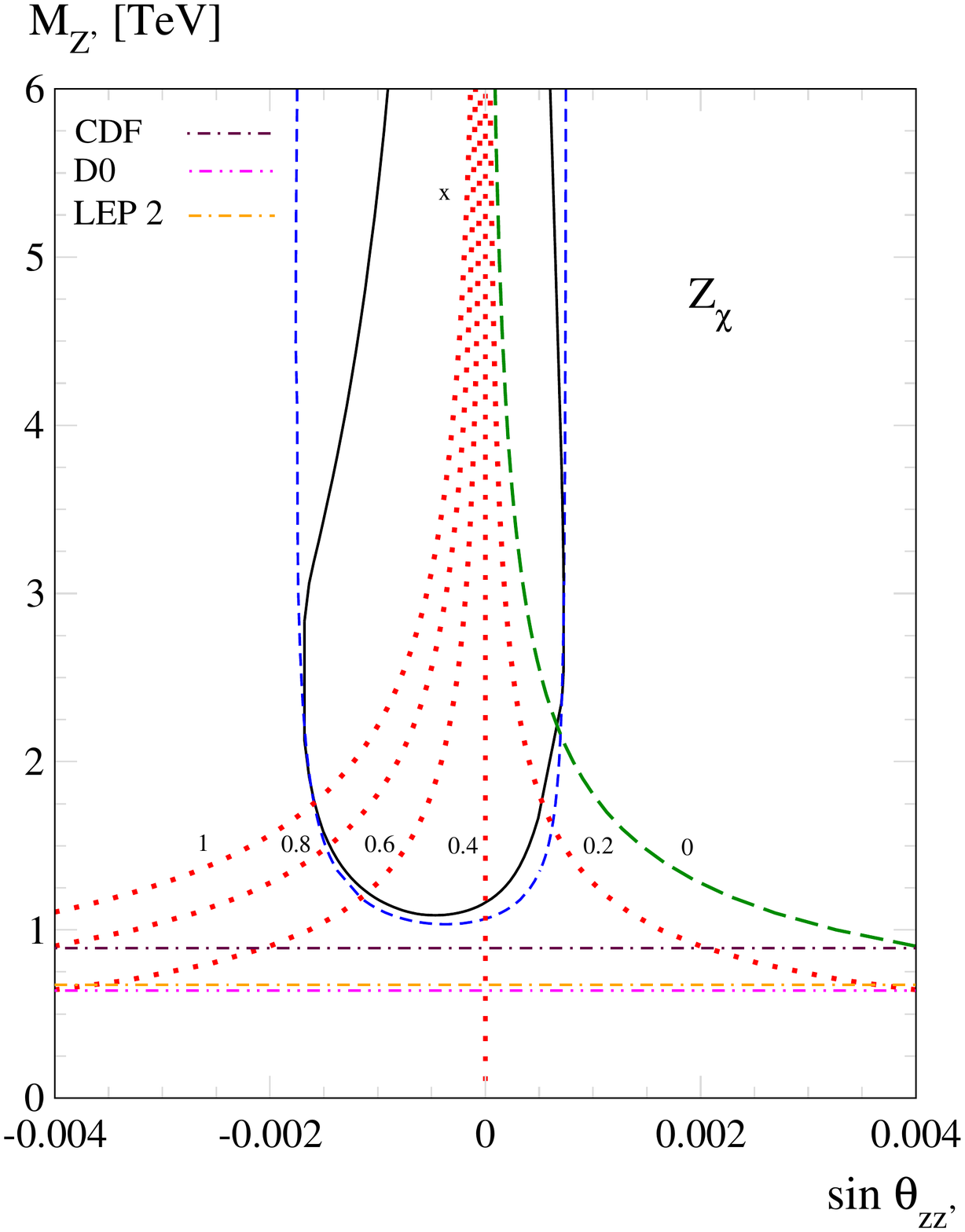}
\end{minipage}
\hspace*{2.5cm}
\begin{minipage}[t]{5cm} 
\includegraphics*[scale=0.34]{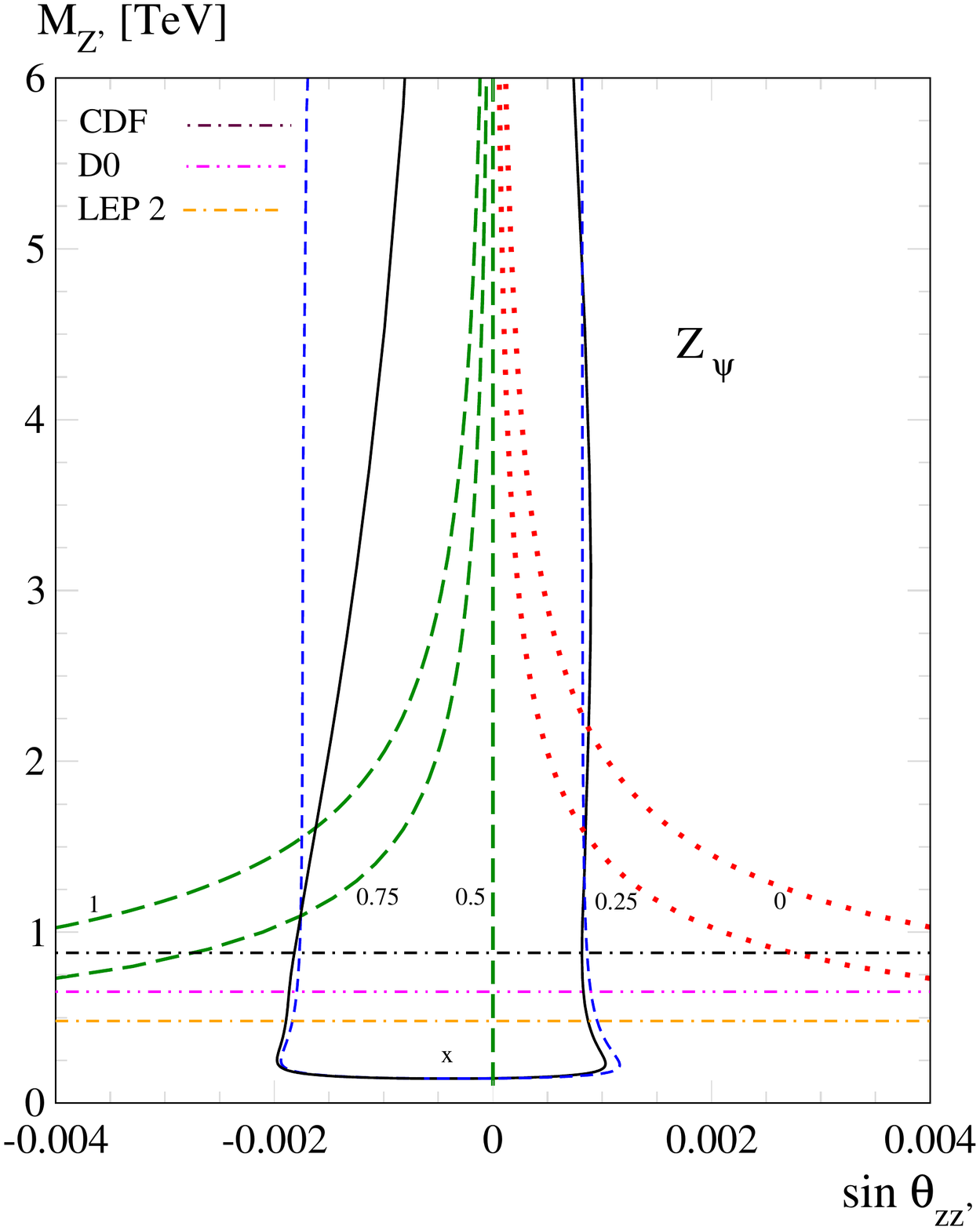}
\end{minipage}
\caption{Experimental constraints on the mass and mixing angle for the $Z_\chi$ 
 and  $Z_\psi$,
 from~\cite{Erler:2009jh}.  The solid lines show the regions allowed by precision electroweak data at 95\% C.L. assuming Higgs doublets and singlets, while the dashed regions allow arbitrary Higgs. The labeled curves assume specific
ratios of Higgs doublet VEVs.}
\label{chipsi}
\end{figure*}

All of these existing limits are listed for a variety of models in Table \ref{table:1}
and the allowed regions in mass and mixing are displayed in Figure \ref{chipsi} for two examples
 in the often studied \zpr\ models ~\cite{Langacker:1984dc,Hewett:1988xc,Langacker:2008yv} based on the
 $E_6$ decomposition  $E_6\ra SO(10)\x U(1)_\psi \ra SU(5)\x U(1)_\chi \x U(1)_\psi$.
%: figure  chi psi limits

There are also significant constraints on \zpr s with family nonuniversal couplings, which lead to 
FCNC when fermion mixing is turned on. Such nonuniversal couplings often occur in string constructions,
or for Kaluza-Klein excitations in extra-dimensional models. Constraints from $K$ decays and mixing, and
from $\mu$ decays and interactions, are usually sufficiently stringent to exclude such effects for the first
two families for a TeV \zpr\ with electroweak couplings~\cite{Langacker:2000ju}. However, the third family could be nonuniversal,
and \zpr-mediated effects could account for possible anomalies in the $B$ system~\cite{Barger:2009qs,Barger:2009eq,He:2006bk,Cheung:2006tm,Baek:2008vr,Mohanta:2008ce,Chang:2009wt,susyfcnc:2009}.

There has recently been considerable discussion of a possible
light \zpr\ in the MeV-GeV range (referred to as a $U$-boson~\cite{Fayet:2006sp,Fayet:2007ua})
which only couples to ordinary matter through  kinetic mixing with the photon.
Such a particle, which is motivated by dark matter considerations~\cite{ArkaniHamed:2008qn}, could have implications for or is constrained by, e.g., 
$g_\mu-2$, $e^+e^-\ra U\gamma\ra e^+e^-\gamma$,  the HyperCP events,
and a variety of laboratory and collider experiments~\cite{Borodatchenkova:2005ct,Pospelov:2008zw,Gopalakrishna:2009yz,Baumgart:2009tn,Reece:2009un,Morrissey:2009ur,Essig:2009nc,Batell:2009yf,Bjorken:2009mm,Batell:2009di,Schuster:2009au,ArkaniHamed:2008qp,Cheung:2009su}.

%: table
\newlength{\tspace}
\setlength{\tspace}{3mm}
\begin{table*}[tb!]
\caption{95\% C.L. limits on \mzp\ and central values and 95\% C.L. upper and lower limits on $\sin \theta_{Z\zpr}$
for a variety of models. The results are updated from~\cite{Erler:2009jh}, where the
models are defined.}
\label{table:1}
%\centering
\begin{tabular}{|c||r|r|r|r||r|r|r||c|} 
\hline
 $Z'$ & \multicolumn{4}{c||}{$M_{Z'}$ [GeV]} & \multicolumn{3}{c||}{$\sin\theta_{Z\zpr}$}  & $\chi^2_{\rm min}$ \\ \hline
& electroweak& CDF & D\O\ & LEP~2 & $\sin\theta_{Z\zpr}$ & $\sin\theta_{Z\zpr}^{\rm min}$ & $\sin\theta_{Z\zpr}^{\rm max}$ & \\ 
\hline\hline 
$Z_\chi$         & \hspace{\tspace}1,141 &    892 & \hspace{\tspace} 800 &    673 & \hspace{\tspace} $-0.0004$ & \hspace{\tspace} $-0.0016$ & \hspace{\tspace} 0.0006 & 47.3 \\ \hline
$Z_\psi$         &    147 &    878 & 763 &    481 & $-0.0005$ & $-0.0018$ & 0.0009 & 46.5 \\ \hline
$Z_\eta$         &    427 &    982 & 810 &    434 & $-0.0015$ & $-0.0047$ & 0.0021 & 47.7 \\ \hline
$Z_I$              & 1,204 &    789 & 692 &            & $ 0.0003$ & $-0.0005$ & 0.0012 & 47.4 \\ \hline
$Z_S$            & 1,257 &    821 & 719 &            & $-0.0003$ & $-0.0013 $& 0.0005 & 47.3 \\ \hline
$Z_N$            &    623 &    861 & 744 &            & $-0.0004$ & $-0.0015$ & 0.0007 & 47.4 \\ \hline
$Z_R$            &    442 &            &         &            & $-0.0003$ & $-0.0015$ & 0.0009 & 46.1 \\ \hline
$Z_{LR}$       &    998 &    630 &         &    804 & $-0.0004$ & $-0.0013$ & 0.0006 & 47.3 \\ \hline
$Z_{\not{L}}$ &   (803) & (740) \hspace{-8pt} & & & $-0.0015$ & $-0.0094$ & 0.0081 & 47.7 \\ \hline
$Z_{SM}$      & 1,403 & \hspace{\tspace} 1,030 & 950 & \hspace{\tspace} 1,787 & $-0.0008$ & $-0.0026$ & 0.0006 & 47.2 \\ \hline
$Z_{string}$  & 1,362 &            &         &            & $ 0.0002$ & $-0.0005$ & 0.0009 & 47.7 \\ \hline\hline
SM                  & \multicolumn{4}{c||}{$\infty$}                               & \multicolumn{3}{c||}{0}                    & 48.5 \\ \hline
\end{tabular}
\end{table*}

\section{The LHC}\label{lhcpotential}
\subsection{discovery}
\begin{figure}[htb!]
\includegraphics*[scale=0.5]{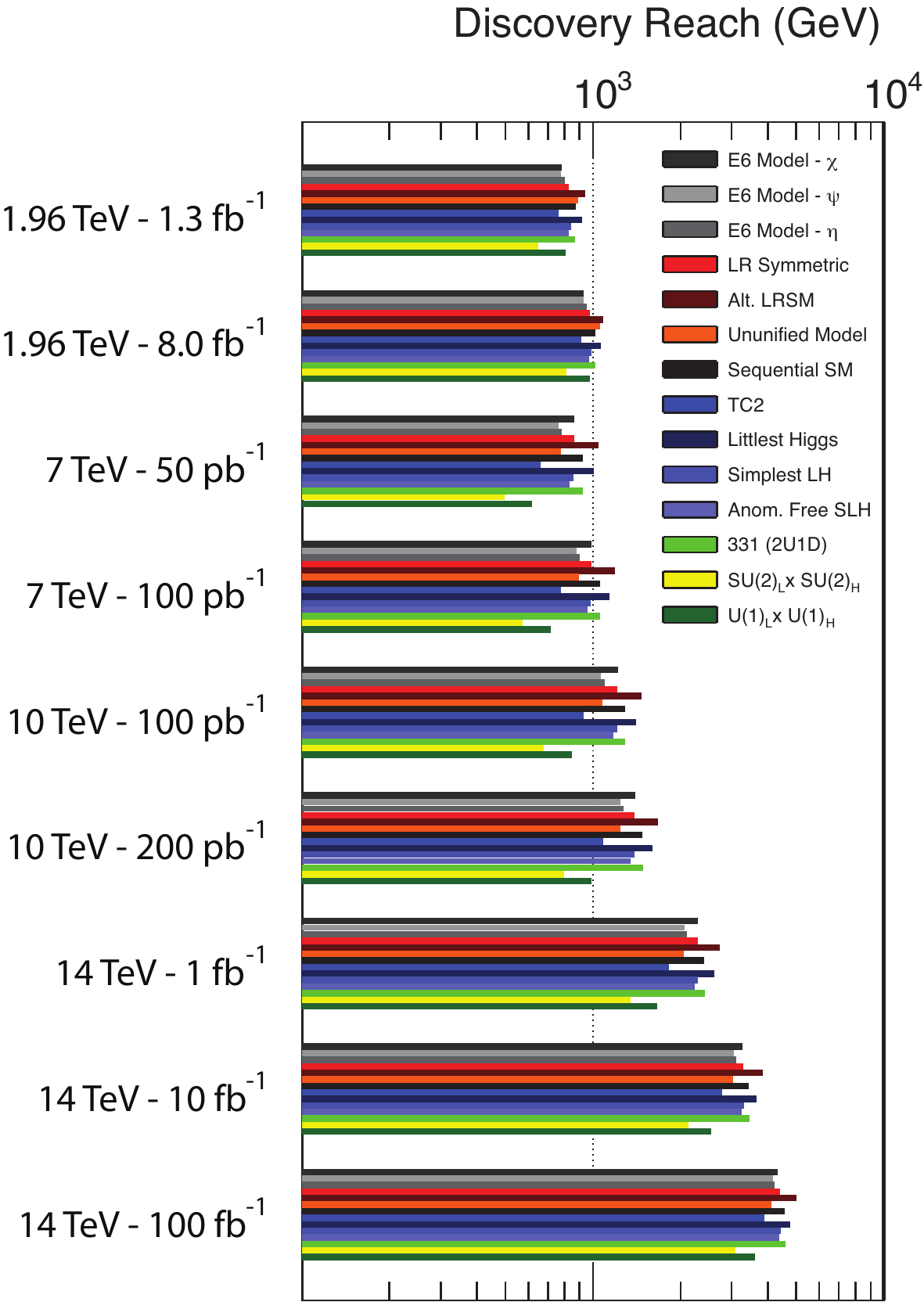}
\caption{LHC discovery reach, based on 5 dilepton events, for typical \zpr\ models as a function of energy and 
integrated luminosity, from \cite{Diener:2009vq}.}\label{discreach}
\end{figure}

The LHC should ultimately have a discovery reach for \zpr s with electroweak-strength couplings to
$u,d,e,$ and $\mu$  up to $\mzp\sim 4-5$ TeV~\cite{Weiglein:2004hn,Dittmar:2003ir,Kang:2004bz,Diener:2009vq}. This is based on decays into $\ell^+\ell^-$ where $\ell=e$ or $\mu$, and  assumes $\sqrt{s}= 14$ TeV and $ \mathcal{L}_I= \int \mathcal{L} dt =100$ fb$^{-1}$.
The reach for a number of models is shown for  various energies and integrated luminosities
in Figure \ref{discreach}. A recent detailed study emphasized the \zpr\ discovery potential in early LHC running at 
 lower energy and luminosity for couplings
 to $B-L$ and $Y$~\cite{Salvioni:2009mt}.
 
\begin{figure*}[htb!]
\begin{minipage}[t]{5.0cm} 
\includegraphics*[scale=0.30]{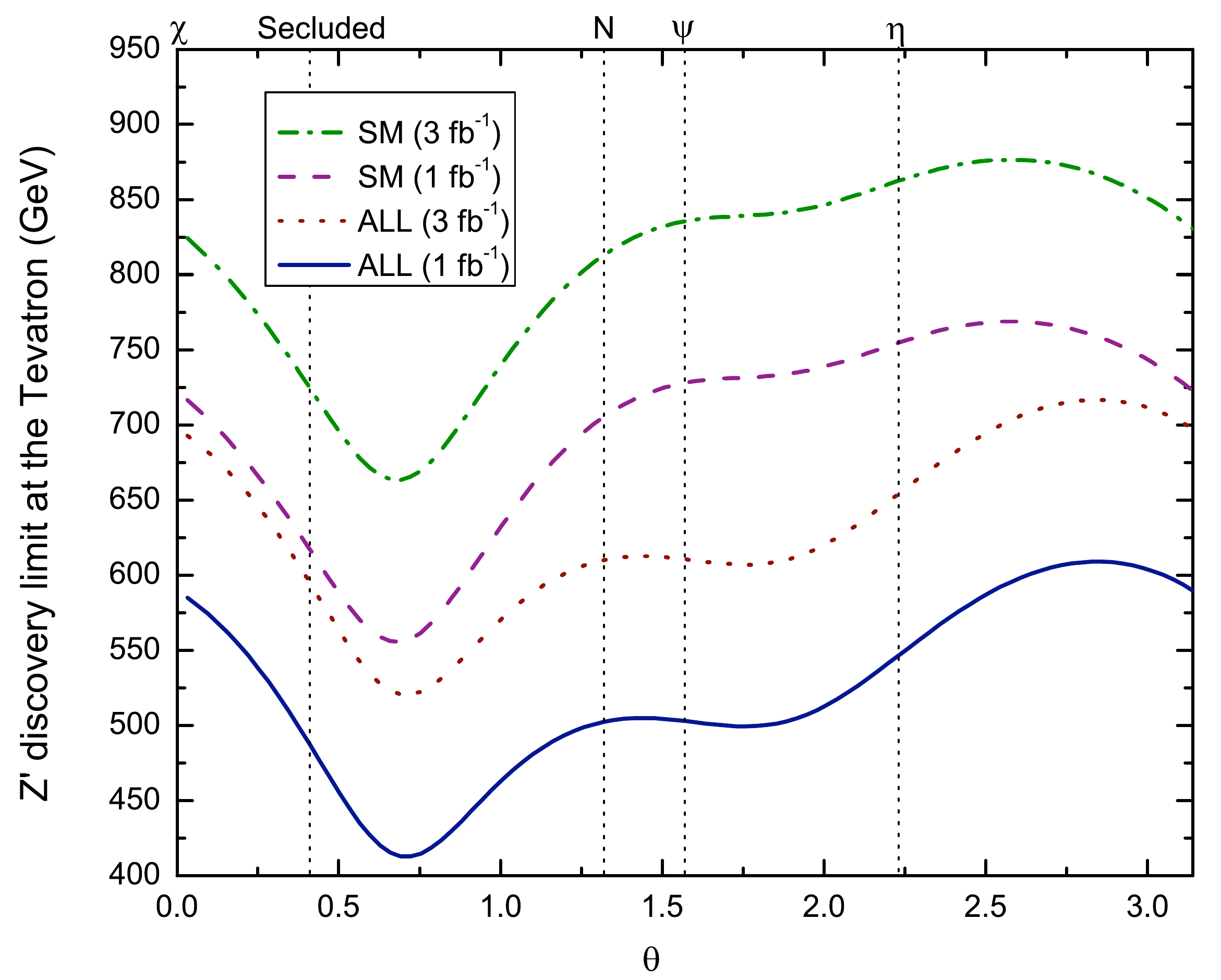}
\end{minipage}
\hspace*{2.5cm}
\begin{minipage}[t]{5.0cm} 
\includegraphics*[scale=0.30]{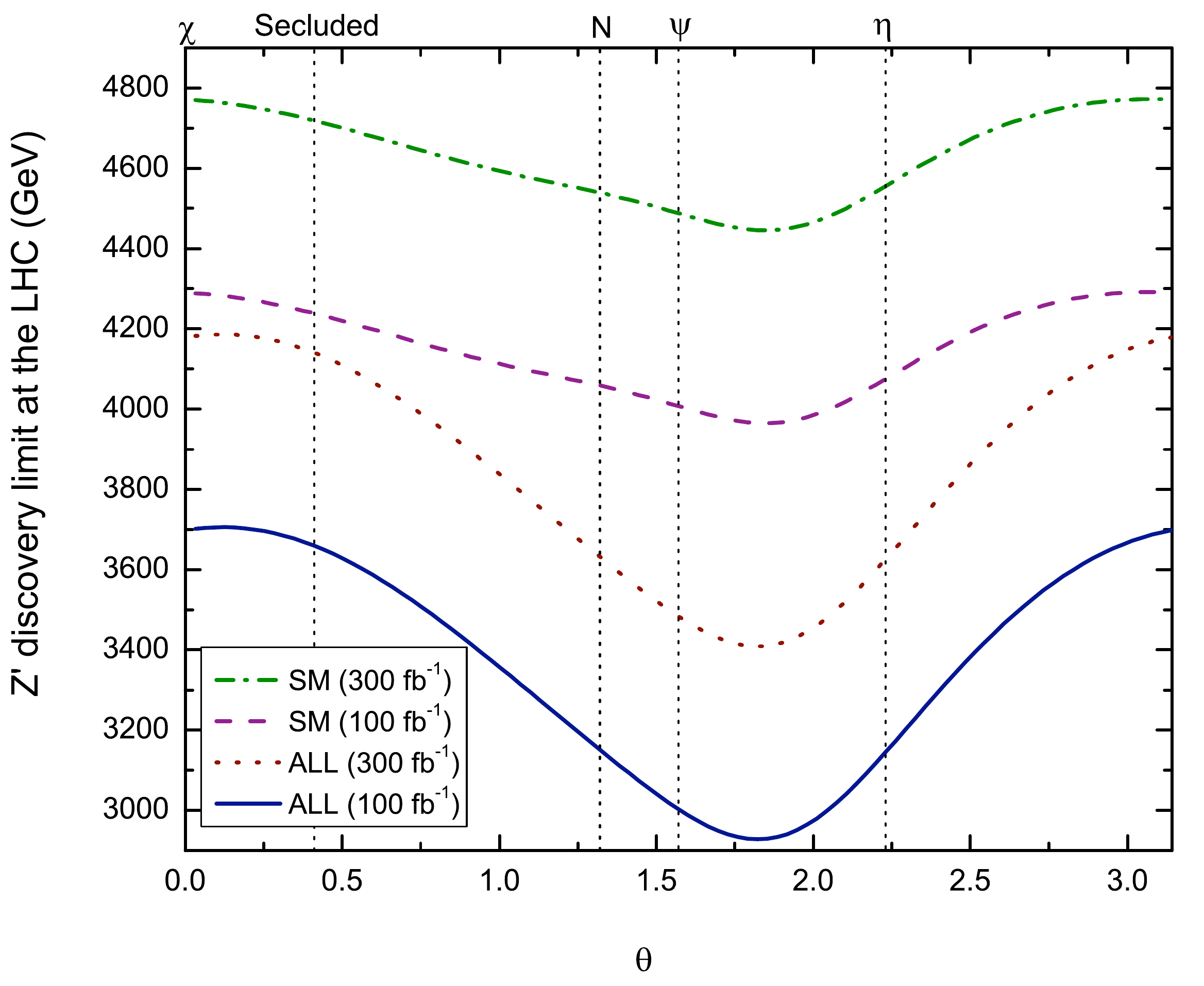}
\end{minipage}
\caption{Discovery reach of the Tevatron and LHC (at 14 TeV) for $E_6$ models,
 assuming decays (a) into SM particles only (SM) and (b) allowing
unsuppressed decays into exotics and sparticles (ALL), based on 10 dilepton events.
The charges are $Q= Q_\chi \cos \theta + Q_\psi \sin \theta$, where $Q_\chi$ and $Q_\psi$ are
associated with $SO(10)$ and $E_6$, respectively. From \cite{Kang:2004bz}.}
\label{widtheffect}
\end{figure*}

The cross section for $pp\ra f\bar f$ (or $ \bar p p \ra  f\bar f$) for a specific final fermion $f$
is just
\beq
\sigma_{\zpr}^f \equiv \sigma_{\zpr} B_f =N_f/ \mathcal{L}_I ,
\eeql{crosf}
where $B_f=\Gamma_f/\Gamma_{Z'}$ is the branching ratio into $f\bar{f}$, 
$\sigma_{\zpr}=\int  \frac{d\sigma_{\zpr}}{dy} dy$,
 and $N_f$ is the number of produced $f\bar f$ pairs for integrated luminosity $ \mathcal{L}_I $.
 For given couplings to the SM particles, $\sigma_{\zpr}^f$ and therefore the discovery reach
 depend on the total width $\Gamma_{Z'}$. For example, 
 in the $E_6$ models
 $\Gamma_{Z'}/\mzp$ can vary from $\sim 0.01-0.05$ depending on whether the important open channels
 include light (compared to \mzp) superpartners and exotics in addition to the SM fermions~\cite{Kang:2004bz}.
 The consequences for the discovery reaches at the Tevatron and LHC are illustrated in
 Figure   \ref{widtheffect}, where it is seen, e.g., that the LHC reach can be reduced by $\sim 1$  TeV if there are many open channels. 
 
There are a number of other potential two-fermion discovery channels,
such as $\tau^+\tau^-$ and $t \bar t $, as mentioned in Section \ref{existing}, while multibody channels will be touched on  in Section \ref{diagnostics}. In principle, the LHC reach in the Drell-Yan dilepton channels can be extended by
using virtual \zpr\ interference effects (cf., the observation of $Z$-propagator effects below the $Z$-pole at TRISTAN~\cite{Mori:1989py}),
though this is difficult in practice~\cite{Rizzo:2009pu}.

%: figure  lhc reach

%: figure  width effect

\subsection{Diagnostics}\label{diagnostics}
The spin of a resonance in the $\ell^+\ell^-$ channel would distinguish a  \zpr\ or other vector from,
e.g., a spin-0 Higgs resonance or a spin-2
Kaluza-Klein graviton excitation. The spin can be determined by the angular distribution in the 
resonance rest frame, which for the spin-1 interactions in \refl{f1} is
\beq
\frac{d \sigma_{\zpr}^f }{d \cos \theta^\ast}  \propto \frac{3}{8} (1+\cos^2\theta^\ast)+A_{FB}^f \cos \theta^\ast,
\eeql{angdist}
where $\theta^\ast$ is the angle between the incident quark  and the $\ell$.
(Magnetic or electric dipole interactions lead to a different distribution~\cite{Chizhov:2009fc})
One does not know which hadron is the source of the $q$
and which the $\bar q$ on an event by event basis, but 
the ambiguity washes out in the determination
of the $1+\cos^2\theta^\ast$ distribution~\cite{Langacker:1984dc,Dittmar:1996my}. 
See~\cite{Osland:2009tn} for a recent detailed study. The \zpr\  spin can also be probed in $t\bar t$ decays~\cite{Barger:2006hm}.

Useful diagnostic probes of the chiral couplings to the quarks, leptons, and other particles, which would help discriminate between \zpr\  models, should be possible for masses up to $\sim$ $2-2.5$ TeV at the LHC, assuming  typical couplings. 
(The gauge coupling $g_2$ can be fixed to the value 
$ g_2=\sqrt{\frac{5}{3}} g \tan \theta_W\sim 0.46$ suggested by some grand unified theories, or alternatively can  be taken as a free parameter if the charges are normalized by some other convention.)

\begin{figure}[htb]
\begin{minipage}[t]{5.0cm} 
\includegraphics*[scale=0.40]{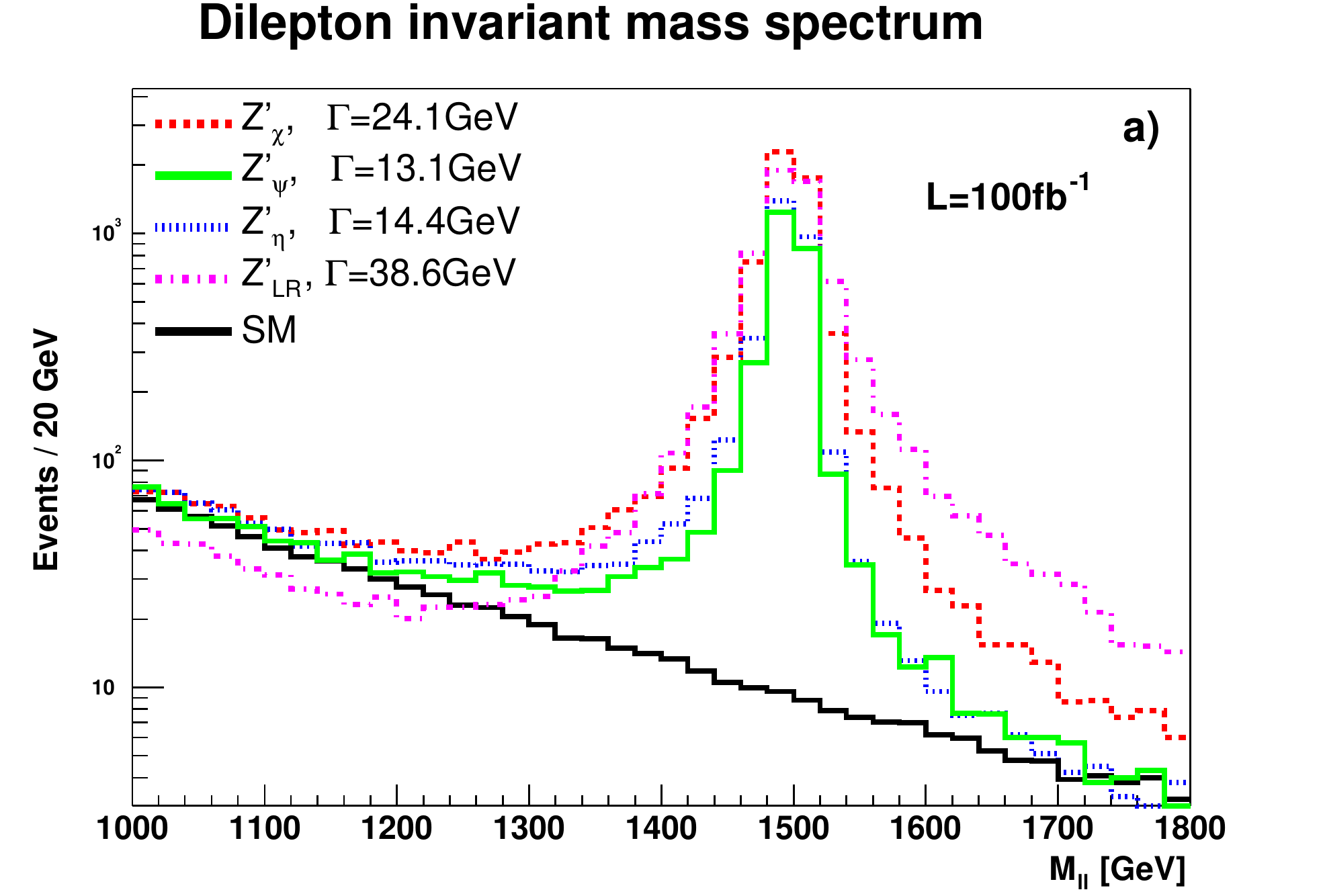}
\end{minipage}
\caption{Dilepton mass spectrum at the LHC
for typical models with $\mzp= 1.5$ TeV, $\sqrt{s}=14$ TeV,  and an integrated luminosity of 100 fb$^{-1}$, from \cite{Dittmar:2003ir}.}
\label{dilepton}
\end{figure}

 For  $pp \to Z'\to \ell^+
\ell^-$ ($\ell=e,\mu$), one would be able to measure
the mass $M_{Z'}$,  the leptonic cross section
 $\sigma_{\zpr}^{ \ell}= \sigma_{\zpr} B_\ell$, and possibly the width $\Gamma_{\zpr}$
 (if it is not too small compared to the detector resolutions). The expected dilepton lineshape
 is illustrated in Figure \ref{dilepton}.
 By itself, $\sigma_{\zpr}^{ \ell}$  is { not}   a useful diagnostic
for the $Z'$ couplings to quarks and leptons:
while $\sigma_{\zpr}$   can be
calculated to within  a few percent for
given $Z'$ couplings,
$B_\ell$
depends strongly on the contribution of exotics
 and sparticles to $\Gamma_{\zpr}$~\cite{Kang:2004bz}.
However, $\sigma_{\zpr}^{ \ell}$
would be a useful indirect probe for the existence of the exotics
 or superpartners.
 The absolute magnitude of the quark and lepton couplings
 is probed by the product
$\sigma_{\zpr}^{ \ell}\Gamma_{\zpr} = 
\sigma_{\zpr} \Gamma_\ell$.
%: figure  dilepton lineshape

\begin{figure*}[htb]
\begin{minipage}[t]{5.0cm} 
\includegraphics*[scale=0.41]{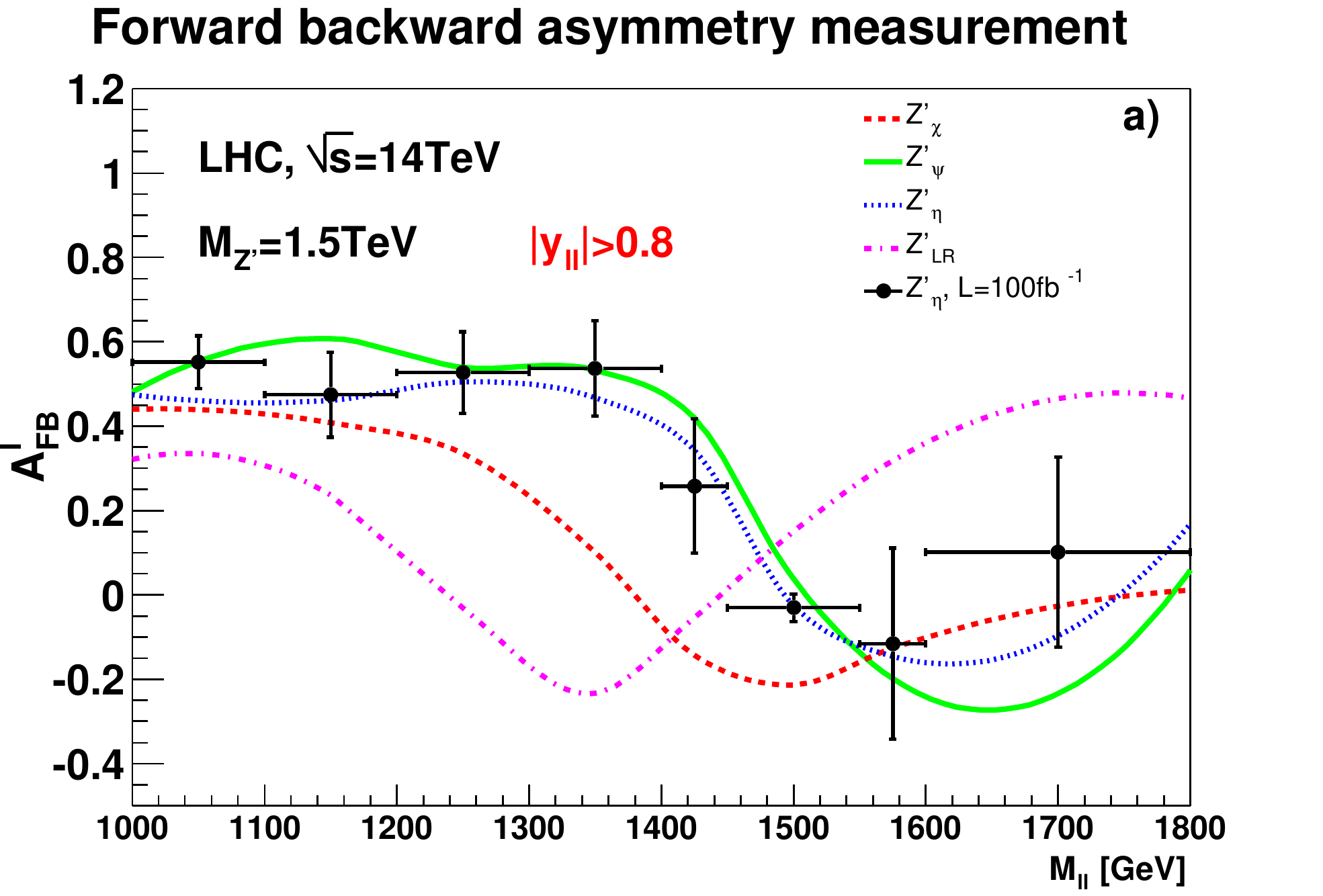}
\end{minipage}
\hspace*{3cm}
\begin{minipage}[t]{5.0cm} 
\includegraphics*[scale=0.41]{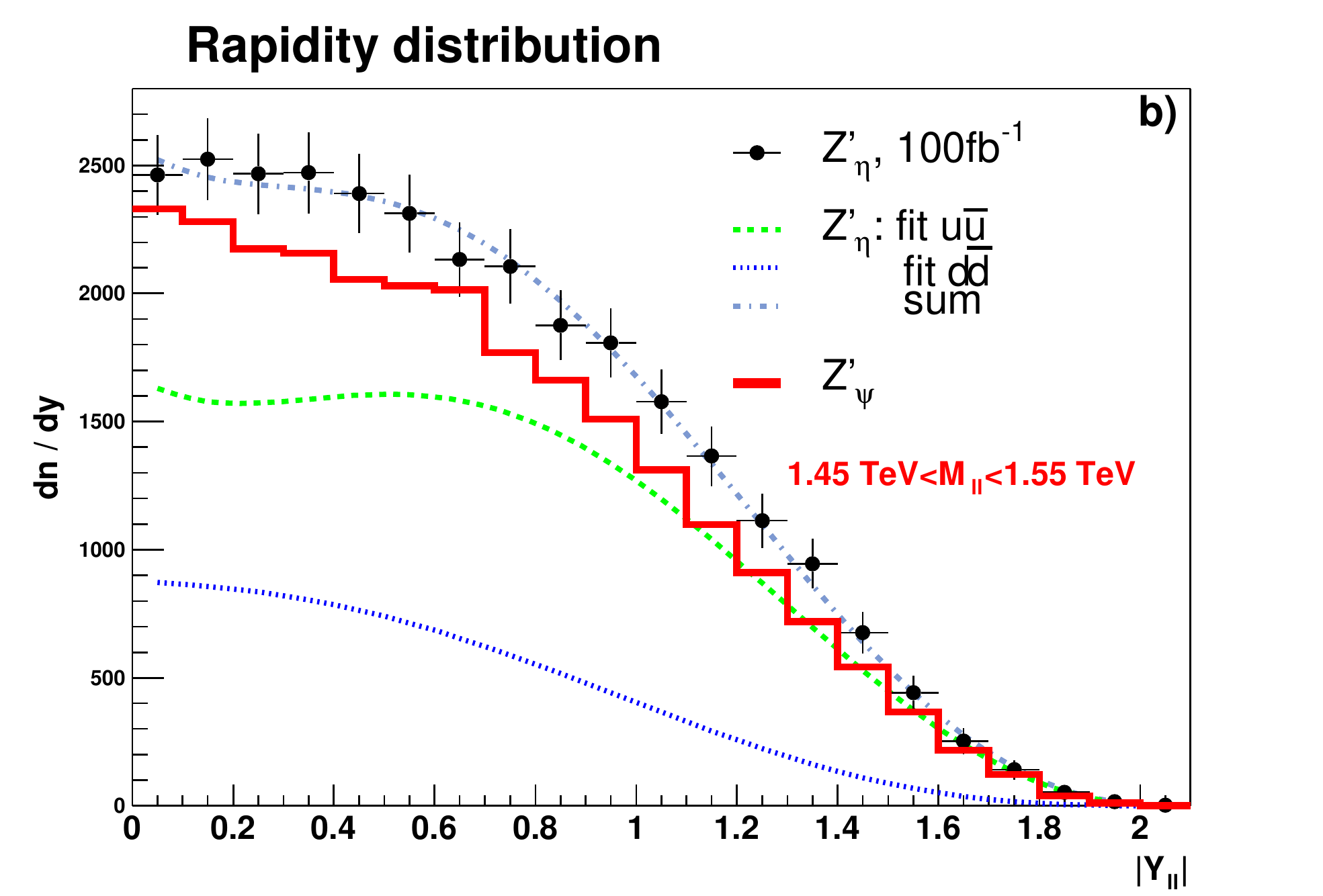}
\end{minipage}
\caption{Forward backward asymmetry and rapidity distributions
for typical models with $\mzp= 1.5$ TeV, $\sqrt{s}=14$ TeV, and an integrated luminosity of 100 fb$^{-1}$, from \cite{Weiglein:2004hn,Dittmar:2003ir}.}
\label{diagnosticdists}
\end{figure*}

The most useful diagnostics   involve the {relative  
strengths} of $Z'$
couplings to ordinary quarks and leptons. The  
forward-backward asymmetry as a function of the $Z'$
rapidity, $A_{FB}^f(y)$~\cite{Langacker:1984dc,Rosner:1995ft,Dittmar:1996my},
avoids the $q\bar q $ ambiguity in {Eq.~\ref{angdist}}.
For $AB\ra\zpr\ra f\bar f $, define $\theta_{CM}$ as the angle of fermion $f$
with respect to the direction of hadron $A$
in the \zpr\ rest frame, and let $F$ ($B$) be the cross section for fixed rapidity $y$
with $\cos \theta_{CM} >0$ ($<0$). Then, $A_{FB}^f(y)\equiv(F-B)/(F+B)$, with 
\begin{align}
&F\pm B  \sim    \left[\begin{array}{c} 4/3 \\ 1\end{array} \right]     \notag   \\
 &\x  \sum_{i} \Bigl( f_{q_i}^A(x_1)f_{\bar q_i}^B(x_2)\pm f_{\bar q_i}^A(x_1)f_{q_i}^B(x_2) \Bigr)\label{abfy}   \\
& \x  \Bigl( \epsilon_L(q_i)^2\pm\epsilon_R(q_i)^2 \Bigr)
 \Bigl( \epsilon_L(f)^2\pm\epsilon_R(f)^2 \Bigr).\notag
\end{align}
 Clearly, $A_{FB}^f(y)$ vanishes for $pp$ at $y=0$,
but can be nonzero at large $y$ where there is more likely a valence $q$ from the first proton
and sea $\bar q$ from the other. The leptonic forward-backward asymmetry is
sensitive to a combination of quark and lepton chiral couplings and is 
a powerful discriminant between models, as can be seen in Figure \ref{diagnosticdists}.
An variant definition of the asymmetry based on the pseudorapidities of the leptons is another
possibility~\cite{Diener:2009ee}.
 
The 
ratio of cross sections  for $\zpr \ra \ell^+\ell^-$ in different rapidity bins~\cite{delAguila:1993ym} gives
information on the relative $u$ and $d$ couplings (Figure \ref{diagnosticdists}). Possible observables in other two-fermion final state channels
include the polarization of produced $\tau$'s~\cite{Anderson:1992jz};
the $pp\to Z'\to j  j$ cross section~\cite{Rizzo:1993ie,Weiglein:2004hn}; and branching ratios,
forward-backward asymmetries, and spin correlations for $b\bar b$ and $t \bar t$~\cite{Barger:2006hm,Godfrey:2008vf,Arai:2008qa,Jung:2009jz}.
There are no current plans for polarization at the LHC, but polarization asymmetries
at a future or upgraded hadron collider would provide another useful diagnostic~\cite{Fiandrino:1992fa}.
Family nonuniversal but flavor conserving effects are discussed in~\cite{Chen:2008za,Salvioni:2009jp}.

 In four-fermion final state  channels the
rare decays $\zpr\ra V f_1 \bar f_2$, where $V=W$ or $Z$ is radiated from the \zpr\ decay products, have a double logarithmic enhancement. 
In particular, $Z'\to W\ell\nu_{\ell}$ (with $W\ra$ hadrons and an  $\ell\nu_{\ell}$ transverse
mass  $> 90$ GeV to separate from SM background) may be observable
and projects out the left-chiral lepton couplings~\cite{Rizzo:1987rw,Cvetic:1991gk,Hewett:1992nf}.
Similarly, the associated productions $pp\rightarrow Z' V$
with $V=(Z,W)$~\cite{Cvetic:1992qv}  and $V=\gamma$~\cite{Rizzo:1992sh}
could yield information on the quark chiral couplings. The processes $pp\ra \zpr Z$ or $ \zpr\gamma$
with the \zpr\ decaying invisibly  into neutrinos or hidden sector particles
may also be observable and could serve as a discovery mode
if the \zpr\ does not couple to charged leptons~\cite{Petriello:2008pu,Gershtein:2008bf}.
The importance of the width for invisible \zpr\ decays for constraining certain extra-dimensional models has been
emphasized in~\cite{Gninenko:2008yq}.

Decays into two bosons, such as $\zpr\ra W^+ W^-, Zh,$ or $W^\pm H^\mp$,
can usually occur only by $Z-\zpr$ mixing or with amplitudes related to the mixing. 
However, this suppression may be compensated for the longitudinal
modes of the $W$ or $Z$ by the large polarization vectors, with components scaling as
$\mzp/M_W$~\cite{Robinett:1981yz,Rizzo:1985kn,Nandi:1986rg,delAguila:1986ad,Barger:1987xw,Baer:1987eb,Gunion:1987jd,Deshpande:1988py}.
For example, $\Gamma(\zpr \ra W^+ W^-)\sim \theta_{Z\zpr} ^2,$ which appears to be hopelessly
small to observe. However, the enhancement factor is $\sim (\mzp/M_W)^4$. Thus, from
{Eq.~\ref{f8}}, these factors compensate, leaving a possibly observable rate that
in principle could give information on the Higgs charges. In the limit of $\mzp \gg M_Z$ one has
\beq
\begin{split}
\Gamma(\zpr \ra W^+ W^-) =& \frac{g_1^2 \theta_{Z\zpr} ^2\mzp}{192\pi} \left(\frac{\mzp}{M_Z}\right)^4\\
=& \frac{g_2^2 C^2 \mzp}{192\pi}.
\end{split}\eeql{zpww}
%\beq
%\Gamma_{W^+ W^-} = \frac{g_1^2 \theta_{Z\zpr} ^2\mzp}{192\pi} \left(\frac{\mzp}{M_Z}\right)^4
%= \frac{g_2^2 C^2 \mzp}{192\pi}.
%\eeql{zpww}
The decay $\zpr\ra Z Z$ has recently been considered~\cite{Keung:2008ve}. The Landau-Yang theorem~\cite{Yang:1950rg} can
be evaded by anomaly-induced or $CP$-violating operators involving a longitudinal $Z$.
The LHC reach of spin-1 resonances associated  with electroweak symmetry breaking and the associated
$\zpr\ra W^+W^-$ or $W' \ra ZW$ decays have been studied in~\cite{Alves:2009aa}, and
more complicated decays such as $\zpr \ra ggg$ or $gg\gamma$  in~\cite{FloresTlalpa:2009jh}.

An alternative source of triple gauge vertices involves anomalous \uprm\ symmetries, which often occur in string constructions.
The anomalies must
be cancelled by a generalized Green-Schwarz mechanism. 
The \zpr\ associated with the \uprm\ acquires a  string-scale mass
by what is essentially the St\"uckelberg mechanism, and effective trilinear vertices may be generated between the \zpr\ and the SM
gauge bosons~\cite{Coriano:2005js,Anastasopoulos:2006cz}. If there are  large extra dimensions
the string scale and therefore the \zpr\ mass may be very low, e.g., at the TeV scale, 
with  anomalous decays into $ZZ$, $WW$, and $Z\gamma$, e.g.,~\cite{Armillis:2008vp,Kumar:2007zza,Anastasopoulos:2008jt}.

Some \zpr\ models lead to distinctive multi-lepton decay modes at a possibly observable rate that are almost entirely free of SM backgrounds.
For example, a \zpr\ could decay into $\ell \bar \ell \ell \bar \ell $ via intermediate sneutrinos
in an $R$-parity violating supersymmetric model~\cite{Lee:2008cn}, or $\zpr\ra 3\ell 3 \bar \ell$ 
by an intermediate $ZH\ra 3Z$ in some models with extended Higgs structures~\cite{Barger:2009xg}.
The latter could occur even in leptophobic models (i.e., with no direct coupling to leptons).
A light (GeV scale) \zpr, suggested by some recent dark matter models, would be highly boosted
at the LHC, leading to  narrow ``lepton jets''
from $Z' \ra \ell^+ \ell^-$ and possible displaced vertices, e.g.,~\cite{ArkaniHamed:2008qp,Cheung:2009su,Langacker:2009im}.

Global studies of the possible LHC diagnostic possibilities for determining ratios
of chiral charges in a model independent way and discriminating models are given in~\cite{Dittmar:2003ir,Petriello:2008zr,Diener:2009vq,Dittmar:1996my,delAguila:1993ym,Cvetic:1995zs,Li:2009xh}. The complementarity of LHC and ILC observations is
especially emphasized in~\cite{Cvetic:1995zs,delAguila:1993rw,delAguila:1995fa,Weiglein:2004hn}.

%:figure fb and rapidity

\section{Other LHC Implications}\label{other}
There are several other implications of a \zpr\ for the LHC. For example, 
TeV scale \uprm\ models generally involve an extended Higgs sector, requiring at least a SM singlet $S$
to break the \uprm\ symmetry.  New $F$ and $D$-term contributions can relax the theoretical upper
limit of $\sim 130$ GeV on the lightest Higgs scalar in the MSSM up to $\sim 150$ GeV, and
smaller values of $\tan \beta$, e.g. $\sim 1$, become possible. Conversely, doublet-singlet
mixing can allow a Higgs lighter than the direct SM and MSSM limits. Such mixing as well as
the extended neutralino sector can lead to non-standard collider signatures, e.g.,~\cite{Accomando:2006ga,Barger:2006dh,Ham:2009bu}.

\uprm\ models also have extended neutralino sectors~\cite{Barger:2007nv,Choi:2006fz}, involving at least the
$\tilde Z'$ gaugino and the $\tilde S$ singlino, allowing  non-standard couplings (e.g., light singlino-dominated),
extended cascades, and modified possibilities for cold dark matter, $g_\mu-2$, etc.

Most \uprm\ models (with the exception of those involving $B-L$ and $Y$) require new exotic fermions to cancel
anomalies. These are usually non-chiral with respect to the SM (to avoid precision electroweak constraints) but chiral under the \uprm. A typical example is a pair of \st-singlet colored quarks $D_{L,R}$  with charge $-1/3$.
Such exotics may decay by mixing, although that is often forbidden by $R$-parity. They may also decay by diquark or leptoquark couplings, or they be quasi-stable, decaying by higher-dimensional operators~\cite{Kang:2007ib,King:2005jy,Athron:2009bs}.

A heavy  \zpr\ may decay efficiently into sparticles, exotics, etc., constituting a ``SUSY factory''~\cite{Kang:2004bz,Lee:2008cn,Baumgart:2006pa,Cohen:2008ni,Ali:2009md}.

 For other theoretical, experimental, and cosmological/astrophysical \zpr\  implications  see~\cite{Langacker:2008yv,Langacker:2009im}.

%\cite{*}
%\bibliographystyle{h-elsevier} 
%\bibliographystyle{href-elsevier} 
%\bibliographystyle{h2-elsevier} 
%:Bibliography

\bibliography{pgl_zp-lhc}
\end{document}